\documentclass[8.5pt,twoside,twocolumn]{article}
\usepackage[super,sort&compress,comma]{natbib} 
\usepackage{mhchem}
\usepackage{times,mathptmx}
\usepackage{sectsty}
\usepackage{balance} 

\usepackage{graphicx} %
\usepackage{lastpage}
\usepackage[format=plain,justification=raggedright,singlelinecheck=false,font=small,labelfont=bf,labelsep=space]{caption} 
\usepackage{fancyhdr}
\begin{document}

\thispagestyle{plain}
\fancypagestyle{plain}{
\renewcommand{\headrulewidth}{1pt}}
\renewcommand{\thefootnote}{\fnsymbol{footnote}}
\renewcommand\footnoterule{\vspace*{1pt}%
\hrule width 3.4in height 0.4pt \vspace*{5pt}} 
\setcounter{secnumdepth}{5}

\makeatletter 
\def\subsubsection{\@startsection{subsubsection}{3}{10pt}{-1.25ex plus -1ex minus -.1ex}{0ex plus 0ex}{\normalsize\bf}} 
\def\paragraph{\@startsection{paragraph}{4}{10pt}{-1.25ex plus -1ex minus -.1ex}{0ex plus 0ex}{\normalsize\textit}} 
\renewcommand\@biblabel[1]{#1}            
\renewcommand\@makefntext[1]%
{\noindent\makebox[0pt][r]{\@thefnmark\,}#1}
\makeatother 
\renewcommand{\figurename}{\small{Fig.}~}
\sectionfont{\large}
\subsectionfont{\normalsize} 

\fancyfoot{}
\fancyfoot[RO]{\footnotesize{\sffamily{1--\pageref{LastPage} ~\textbar  \hspace{2pt}\thepage}}}
\fancyfoot[LE]{\footnotesize{\sffamily{\thepage~\textbar\hspace{3.45cm} 1--\pageref{LastPage}}}}
\fancyhead{}
\renewcommand{\headrulewidth}{1pt} 
\renewcommand{\footrulewidth}{1pt}
\setlength{\arrayrulewidth}{1pt}
\setlength{\columnsep}{6.5mm}
\setlength\bibsep{1pt}

\twocolumn[
  \begin{@twocolumnfalse}
\noindent\LARGE{\textbf{Chiral and chemical oscillations in a simple dimerization model$^\dag$}}
\vspace{0.6cm}

\noindent\large{\textbf{Michael Stich,$^{\ast}$\textit{$^{a}$} Celia
    Blanco,\textit{$^{a}$} and David Hochberg\textit{$^{a}$}}}\vspace{0.5cm}

\noindent\textit{\small{\textbf{Received Xth XXXXXXXXXX 201X, Accepted Xth XXXXXXXXX 201X\newline
First published on the web Xth XXXXXXXXXX 201X}}}

\noindent \textbf{\small{DOI: 10.1039/b000000x}}
\vspace{0.6cm}

\noindent \normalsize{We consider the APED model
  (activation-polymerization-epimerization-depolymerization) for
  describing the emergence of chiral solutions within a non-catalytic
  framework for chiral polymerization. The minimal APED model for
  dimerization can lead to the spontaneous appearance of chiral
  oscillations and we describe in detail the nature of these
  oscillations in the enantiomeric excess, and which are the
  consequence of oscillations of the concentrations of the associated
  chemical species.}
\vspace{0.5cm}
 \end{@twocolumnfalse}
  ]

\footnotetext{\dag~Electronic Supplementary Information (ESI) available: [details of any supplementary information available should be included here]. See DOI: 10.1039/b000000x/}

\footnotetext{\textit{$^{a}$~Centro de Astrobiolog\'ia (CSIC-INTA),
    Ctra de Ajalvir km 4, 
    28850 Torrej\'on de Ardoz (Madrid), 
    Spain. Fax: +34 91 520-1074; Tel: +34 91 520-6409; E-mail: stich@cab.inta-csic.es}}

\section{Introduction}
\label{sec:intro}

Chemical oscillations are a paradigmatic example of temporal
self-organization in non-equilibrium systems and have been at the
center of intense experimental and theoretical research for
decades~\cite{Kuramoto84,Scott94,Kapral95}, with theoretical work 
going back at least to the times of Lotka~\cite{LotkaJPC10}.  
At the heart of these
spontaneous oscillations are a positive feedback loop, often an
autocatalytic step, and a negative control loop, responsible for the
saturation of the growth process and which operates often on a slower
time scale.

The mathematical description of chemical oscillations relies usually
on ordinary differential equations readily derived from the chemical
rate laws. The number of reacting species determines the number of
variables and chemical reaction kinetics implies that these equations
are nonlinear.  The number, nature, and stability of the solutions of
the system are studied through standard methods from dynamical systems
theory, bifurcation theory, among
others~\cite{Guckenheimer83,Kuznetsov95}. One simple possibility to
obtain stable oscillations is the Hopf bifurcation scenario, discussed
in more detail below.

Instabilities observed in chemical systems are nowadays
well-established in theory and experiment, and in tradition of the
famous Belousov-Zhabotinsky reaction~\cite{ZaikinN70}, most
experiments refer to chemical reactions in solution. Nevertheless,
surface chemical reactions and an increasing number of biochemical
reactions have also been studied
extensively~\cite{Murray89,Walgraef97}.

In this article, we study a model purporting to describe general
chiral polymerization processes in a prebiotically relevant chemical
system, devised for explaining the appearance of
homochirality~\cite{PlassonPNAS04}. Although there are many different
theoretical (and experimental) models tackling the open question of
the origin of biological homochirality, i.e., the preference of living
matter for only one of two chiral states of an otherwise chemically
identical molecule, at the heart of many of these approaches lies a
mechanism responsible for spontaneous mirror symmetry breaking. The
initially small chiral fluctuation must then be amplified to a state
that is useful for biotic evolution. A precondition for this to happen
is that the chirality must be preserved and transmitted to the rest of
the system.

In this context, the implications of chiral oscillations for chirality
transmission are far-reaching: indeed, the memory of the sign of the
initial primordial chiral fluctuation is erased and diluted as it were
by whatever subsequent oscillations are present in the enantiomeric
excess. This phenomenon, if and when it occurs in any model purporting
to have relevance to prebiotic chemistry, adds a further element of
uncertainty and stochasticity to the determination of the sign of the
chirality that is \textit{finally} transmitted to the system (say,
within a closed reacting system with necessarily damped chiral
oscillations). Recently in fact, numerical evidence for such damped
oscillations was found in a chiral polymerization model closed to
matter and energy flow, where the amplitude of the oscillation depends
on the length of the homochiral chain
formed~\cite{BlancoPCCP11}. While such models are successful in
fitting to real chemical data on e.g., relative abundances of chiral
polymers~\cite{BlancoPCCP12}, they usually involve dozens or even
hundreds of nonlinear coupled differential equations, and they do not
lend themselves easily to a systematic controlled mathematical
analysis of the kind needed to study rigorously oscillatory phenomena
in detail.

For this essentially technical reason, a detailed study of chiral
oscillatory phenomena is best carried out in a simple model, ideally
not too far removed, if possible, in spirit from a more complex "real"
model. Linear stability analysis proves that there is no possibility
for oscillations in the original Frank model, nor within any of its
minimal extensions~\cite{RiboPLA08}. On the other hand, the model
proposed by Plasson et al.~\cite{PlassonPNAS04} is amenable to
theoretical analysis, and it exhibits different stable stationary
states, together with instabilities and even temporal
oscillations. These oscillations refer not only to the concentrations
of the different species, and therefore represent chemical
oscillations, but also to oscillations in the enantiomeric excess of
the system. These chiral oscillations are the focus of this article.

This article is organized as follows: In Sec.~\ref{sec:model} we
introduce the minimal APED model and perform a basic bifurcation
analysis. In Sec.~\ref{sec:bif}, we discuss in more detail the basic
bifurcations observed in the model and display bifurcation diagrams
for other parameters. One of the interesting results is that chiral
oscillations can be expected for parameter values that are chemically
more realistic, at least within simple models based on Frank's scheme,
when these are proposed as the fundamental candidate reaction network
for explaining the salient features of the Soai
reaction~\cite{CrusatsCPC09}.  Finally, we close the article with a
discussion of the results in Sec.~\ref{sec:disc}.

\section{The APED model}
\label{sec:model}

We consider the so-called APED
(activation-polymerization-epimerization-depolymerization) model,
introduced by Plasson et al.~\cite{PlassonPNAS04}, used to describe
the spontaneous onset of homochiral states in a symmetric system of
interacting polymerization products.  The reaction scheme for the
minimal APED dimerization model is represented by the following
chemical transformations
\begin{equation}
\begin{split}
\textrm{L} \stackrel{a}{\longrightarrow}  \textrm{L}^\ast \qquad
\textrm{LL} &\stackrel{h}{\longrightarrow}  \textrm{L}+\textrm{L} \\
\textrm{D} \stackrel{a}{\longrightarrow}  \textrm{D}^\ast \qquad
\textrm{DD} &\stackrel{h}{\longrightarrow}  \textrm{D}+\textrm{D} \\
\textrm{L}^\ast \stackrel{b}{\longrightarrow}  \textrm{L} \qquad
\textrm{LD} &\stackrel{\beta h}{\longrightarrow}  \textrm{L}+\textrm{D} \\
\textrm{D}^\ast \stackrel{b}{\longrightarrow}  \textrm{D} \qquad
\textrm{DL} &\stackrel{\beta h}{\longrightarrow}  \textrm{L}+\textrm{D} \\
\textrm{L} +  \textrm{L}^\ast \stackrel{p}{\longrightarrow}  \textrm{LL} \qquad 
\textrm{LL} &\stackrel{\gamma e}{\longrightarrow}  \textrm{DL} \\
\textrm{D} +  \textrm{D}^\ast \stackrel{p}{\longrightarrow}  \textrm{DD} \qquad
\textrm{DD} &\stackrel{\gamma e}{\longrightarrow}  \textrm{LD} \\
\textrm{L} +  \textrm{D}^\ast \stackrel{\alpha p}{\longrightarrow}  \textrm{DL}\qquad 
\textrm{LD} &\stackrel{e}{\longrightarrow}  \textrm{DD} \\
\textrm{D} +  \textrm{L}^\ast \stackrel{\alpha p}{\longrightarrow}  \textrm{LD} \qquad
\textrm{DL} &\stackrel{e}{\longrightarrow}  \textrm{LL} 
\end{split}
\end{equation}

The model contains deactivated monomers L and D, activated monomers
L$^\ast$ and D$^\ast$, homodimers LL and DD, and heterodimers LD and
DL. The reactions include activation (rate $a$), deactivation (rate
$b$), homochiral polymerization (rate $p$), heterochiral
polymerization (rate $\alpha p$), homochiral hydrolysis (rate $h$),
heterochiral hydrolysis (rate $\beta h$), homochiral epimerization
(rate $e$), and heterochiral epimerization (rate $\gamma e$). In
general, polymerization, hydrolysis, and epimerization processes are
stereoselective, i.e., $\alpha \ne 1$, $\beta \ne 1$, $\gamma \ne
1$. Mass is conserved and therefore the total concentration in
residues
$c=[\mathrm{L}]+[\mathrm{D}]+[\mathrm{L^\ast}]+[\mathrm{D^\ast}]+
2([\mathrm{LL}]+[\mathrm{DD}]+[\mathrm{LD}]+[\mathrm{DL}])$, is
constant.

The chemical reactions transcribe into the following set of ordinary
differential equations:
\begin{subequations}
\begin{align}
\dot{\textrm{[L]}} &=-a\textrm{[L]}
-p\textrm{[L]}\textrm{[L}^\ast\textrm{]} -\alpha
p\textrm{[L]}\textrm{[D}^\ast\textrm{]} + 2h\textrm{[LL]} + \nonumber \\
& \quad + b\textrm{[L}^\ast\textrm{]} +\beta h\textrm{[DL]} + \beta h \textrm{[LD]}\\
\dot{\textrm{[D]}} &=-a\textrm{[D]}
-p\textrm{[D]}\textrm{[D}^\ast\textrm{]} -\alpha
p\textrm{[D]}\textrm{[L}^\ast\textrm{]} + 2h\textrm{[DD]} + \nonumber \\ 
& \quad + b\textrm{[D}^\ast\textrm{]} + \beta h\textrm{[DL]} + \beta h \textrm{[LD]}\\
\dot{\textrm{[L}^\ast\textrm{]}} &=a\textrm{[L]} -p\textrm{[L]}\textrm{[L}^\ast\textrm{]} - \alpha p\textrm{[L}^\ast\textrm{]}\textrm{[D]} - b\textrm{[L}^\ast\textrm{]} \\
\dot{\textrm{[D}^\ast\textrm{]}} &=a\textrm{[D]} -p\textrm{[D]}\textrm{[D}^\ast\textrm{]} - \alpha p\textrm{[D}^\ast\textrm{]}\textrm{[L]} - b\textrm{[D}^\ast\textrm{]} \\
\dot{\textrm{[LL]}} &=p\textrm{[L]}\textrm{[L}^\ast\textrm{]} -h\textrm{[LL]} +e\textrm{[DL]} -\gamma e\textrm{[LL]} \\
\dot{\textrm{[DD]}} &=p\textrm{[D]}\textrm{[D}^\ast\textrm{]} -h\textrm{[DD]} +e\textrm{[LD]} -\gamma e\textrm{[DD]} \\
\dot{\textrm{[DL]}} &=\alpha p\textrm{[L]}\textrm{[D}^\ast\textrm{]} -e\textrm{[DL]} -\beta h\textrm{[DL]} + \gamma e\textrm{[LL]} \\
\textrm{[LD]}
&=(c-\textrm{[L]}-\textrm{[D]}-\textrm{[L}^\ast\textrm{]}-\textrm{[D}^\ast\textrm{]}-2\textrm{[LL]}-\nonumber \\
& \quad -2\textrm{[DD]}-2\textrm{[DL]})/2
\end{align}
\end{subequations}
where due to mass conservation, the concentration of one of the
species, here chosen to be LD, is given by the others.  An important
quantity is the enantiomeric excess in the system. Unless otherwise
stated, we consider the total enantiomeric excess,
$ee=(\textrm{[L]}+\textrm{[L}^\ast\textrm{]}-\textrm{[D]}-\textrm{[D}^\ast\textrm{]}+2\textrm{[LL]}-2\textrm{[DD]})/c$.

\begin{figure*} 
 \centering 
\includegraphics[width=16cm]{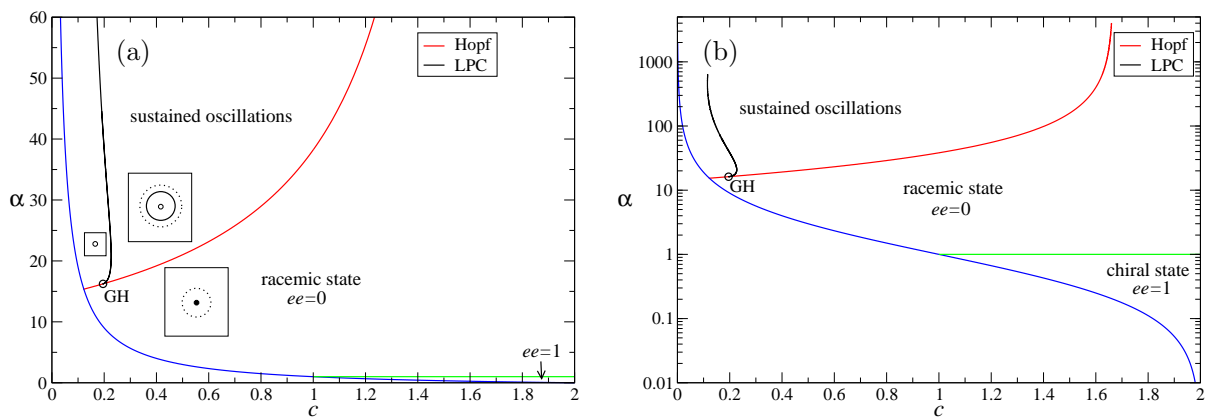}
\caption{Bifurcation analysis of the APED model in the $\alpha-c$
  space for the parameter values $a=h=e=p=1$ and $b=\beta=\gamma=0$.
  The GH point (open circle) is located at GH$=(0.19559,16.23713)$.
  Stable oscillations of the variables (and enantiomeric excess) are
  found in the region between the supercritical Hopf curve and the LPC 
  (for: Limit Point bifurcation of Cycles)
  curve. In (a), $\alpha$ is shown on a linear scale from $0$ to $60$
  and the insets qualitatively display the main solutions: open
  (filled) circle for the unstable (stable) fixed point, i.e., racemic
  state, and solid (dashed) curves for the stable (unstable) limit
  cycle, i.e., oscillations. (b) Same curves shown on a logarithmic
  scale for $\alpha$, for comparison with Fig.~3 of
  Ref.~\cite{PlassonPNAS04}. Note that oscillations cannot be found
  for any $\alpha$ if $c$ becomes sufficiently large. More information
  see text.}
 \label{calpha1}%
\end{figure*}
It is straightforward to search for different stable states within the
system. Following the analysis published in Ref.~\cite{PlassonPNAS04},
we present in Fig.~\ref{calpha1} a bifurcation diagram obtained with
{MatCont}~\cite{DhoggeACMTOMS03} for the $\alpha -c$ space. The
different behaviors of the system depend on the total concentration
$c$ and the relative strength of the hetero- vs. homo-dimerization,
$\alpha$. We start our analysis at the point $c=1$, $\alpha=10$ (other
parameters are $a=h=e=p=1$ and $b=\beta=\gamma=0$), where we find a
stable racemic state (vanishing enantiomeric excess) characterized by
constant and nonvanishing values for all variables (but with [D]=[L],
[DD]=[LL] etc.). We keep $a=1$, $b=0$ for the remainder of this
article.

If we decrease the total matter in the system (lowering $c$), the
system crosses the blue curve and the only stable state is the ``dead
state'', characterized by vanishing concentrations for all species
except for the activated species L$^\ast$ and D$^\ast$. This curve is
given by the equation $c=2a/(p+p\alpha)$, see
Ref.~\cite{PlassonPNAS04}.  If we decrease $\alpha$ while $c>1$, the
system undergoes a pitchfork bifurcation at $\alpha=1$, at which the
racemic state becomes unstable and stable chiral states appear. These
chiral states were the focus of the article~\cite{PlassonPNAS04}, and
therefore we do not discuss them in detail here.

\begin{figure}%
 \centering 
\includegraphics[width=8cm]{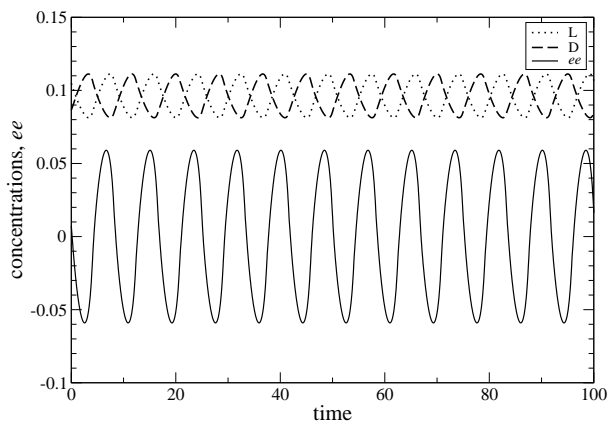}
\caption{Stable oscillations of the concentrations and the
  enantiomeric excess $ee$ observed in the APED model for the
  parameter values $a=h=e=p=c=1$, $b=\beta=\gamma=0$, and $\alpha=40$.
  All concentrations oscillate, for simplicity we just display [L] and
  [D], and have omitted the initial transient. Note that [L] and [D]
  oscillate out of phase.}
 \label{times1}%
\end{figure}
As we increase $\alpha$ for fixed $c$ (which is allowed to have values
as small as $c\approx 0.2$), we observe the onset of oscillations as
the racemic stationary state becomes unstable through a supercritical
Hopf bifurcation (red curve). This means that close to the
bifurcation, oscillations have small amplitude and finite
frequency. These oscillations are exemplified in Fig.~\ref{times1} for
$\alpha=40$, $c=1$, i.e., already at some distance from the Hopf point
located at $\alpha=\alpha_H=38.24$. All concentrations oscillate (with
a period $T=6.52355$), with the chiral species oscillating in
antiphase, and the enantiomeric excess oscillates with the same
period. The range where stable chiral oscillations are observed does
not extend to the hyperbola limiting the dead state.  As $c$ is
decreased for a fixed $\alpha$, the stable limit cycle undergoes a
saddle-node bifurcation of limit cycles (black curve, denoted as LPC,
for: Limit Point bifurcation of Cycles). This means that the stable
limit cycles collides with an unstable one and both disappear (and
hence no limit cycles are present to the left of the black curve).
The fixed point describing the racemic state remains unstable in that
parameter region and numerical simulations (not shown here) indicate
that the system dynamics is governed by irregular oscillations that do
not saturate and finally lead to negative values for the
concentrations, hence not representing any realistic chemical state.

The curve of the supercritical Hopf bifurcation and the curve of
saddle-node bifurcation of limit cycles meet at a Generalized Hopf
(GH) point at GH$=(0.19559,16.23713)$.  This implies that the
supercritical Hopf bifurcation converts into a subcritical one. Hence,
the red curve between the GH point and the point where the Hopf
bifurcation meets the hyperbola $c=2/(1+\alpha)$ represents a
subcritical Hopf bifurcation where the unstable fixed point (above the
curve) becomes stable (below the curve), representing the stable
racemic state, while at the same time an unstable limit cycle
appears. This unstable limit cycle is the one which disappears at the
LPC curve.  The insets of Fig.~\ref{calpha1}(a) show qualitatively the
stationary racemic state and the limit cycles as one moves around the
GH point in $\alpha-c$ parameter space. In the parameter regime where
stable chiral oscillations are observed, another, unstable, limit
cycle is present. For this parameter set, the minimal $\alpha$-value
for oscillations is given by $\alpha_c = 16.237$.

In Fig.~\ref{calpha1}(b), we show the same bifurcation diagram in a
log-linear plot. Contrary to what was displayed in Fig.~3 of
Ref.~\cite{PlassonPNAS04}, the Hopf curve bends upward for $c \to 2$,
demonstrating that the oscillations are not posible if the total
amount of residues is large. Therefore, the region for $c$ for which
stable oscillations are found is bound from below and above.

\section{Bifurcation analysis}
\label{sec:bif}

The bifurcation analysis carried out in Sec.~\ref{sec:model} shows
that for the parameter values $a=h=e=p=1$ and $b=\beta=\gamma=0$,
stable oscillations are only to be found for $\alpha > \alpha_c =
16.237$.  Here, we consider other (obviously not all) regions of the
overall 9-parameter space (we always keep $a=1$, $b=0$) to check how
general is the appearance of chiral oscillations. Since we focus on
the oscillations, we do not show a complete bifurcation diagram.

\begin{figure}%
 \centering 
\includegraphics[width=8cm]{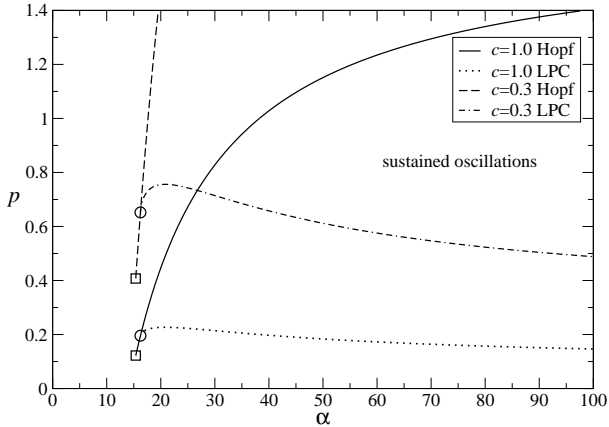}
\caption{Bifurcation analysis in $p-\alpha$ space for the parameter
  values $c=1$ and $c=0.3$.  The area with sustained oscillations is
  found to the right of the Hopf curves and above the LPC curves,
  respectively. The GH and the intersection with $c=2a/(p+p\alpha)$
  (appearance of the dead state) are denoted by open circles and
  squares, respectively. For $c=1$: GH$=(16.23713,0.19559)$,
  intersection at $(15.36557,0.12221)$. For $c=0.3$:
  GH$=(16.23713,0.65197)$, intersection at $(15.36451,0.40739)$.
  Other parameters are $a=h=e=1$ and $b=\beta=\gamma=0$.}
 \label{palpha1}%
\end{figure}
In Fig.~\ref{palpha1} we investigate for two different total
concentrations $c=1$ and $c=0.3$, the region of stable oscillations in
the $p-\alpha$ space. We hence probe at the same time the absolute
influence of the polymerization rate ($p$) and the relative influence
of its stereoselectivity ($\alpha$).  The region of stable
oscillations is again limited by a Hopf curve and an LPC curve. The
two curves are qualitatively similar for both values of $c$, so we
restrict the description to the case $c=1$. If we choose, e.g.,
$p=1$, and increase $\alpha$, the system undergoes a supercritical
Hopf bifurcation at $\alpha_H=38.24$ as already seen above. Hence, to
the right of the Hopf curve, the region of stable oscillations is
found, which is limited from below by the LPC curve. This means that
if $p$ takes smaller values, smaller $\alpha$'s are necessary to
induce oscillations, although there is a minimum value for $p$, below
which no oscillations can be found, here, $p\approx 0.2$,
independently of $\alpha$. The LPC and Hopf curves merge at the GH
point, giving again a minimum value for $\alpha=\alpha_c$, below
which no oscillations are observed. For smaller $c$, larger $p$ have
to be chosen to obtain stable oscillations.
\begin{figure*}
 \centering 
\includegraphics[width=18cm]{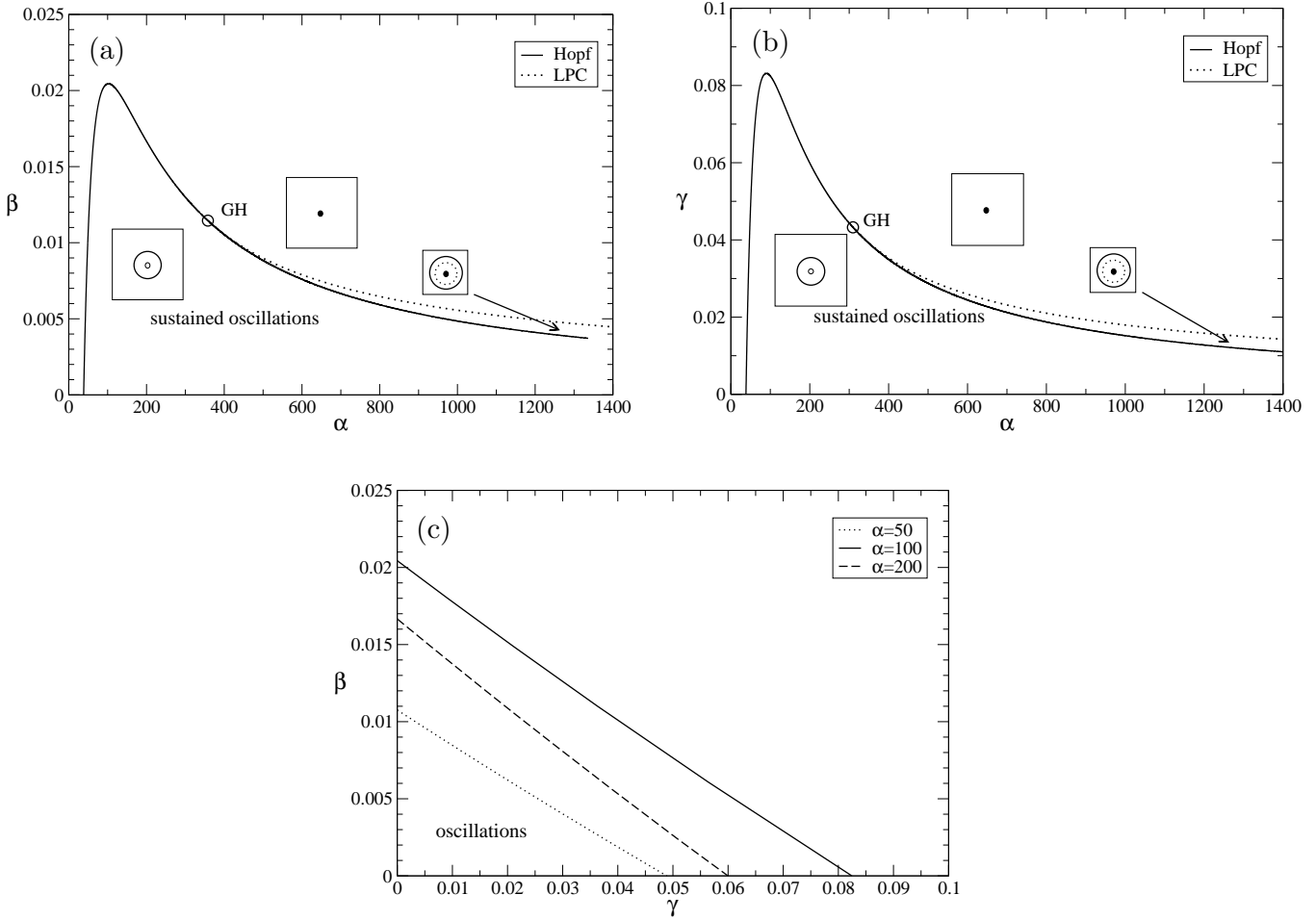}
\caption{(a) Bifurcation analysis in $\beta-\alpha$ space.  The area
  with sustained oscillations is found below the supercritical Hopf
  curve (to the left of GH) and below the LPC curve (to the right of
  GH). The insets qualitatively display the main solutions: filled
  (open) circle for the stable (unstable) fixed point, i.e., racemic
  state, and solid (dashed) curves for the stable (unstable) limit
  cycle, i.e., oscillations. Note that the bifurcation scenario is
  slightly different than in Fig.~\ref{calpha1}, giving rise to a thin
  parameter region where bistability among the stable fixed point and
  the stable limit cycle exists. The GH point is located at
  GH$=(358.033236,0.011450)$, Other parameters are $a=c=h=p=e=1$ and
  $b=\gamma=0$. (b) Bifurcation analysis in $\gamma-\alpha$ space. The
  area with sustained oscillations is found below the supercritical
  Hopf curve (to the left of GH) and below the LPC curve (to the right
  of GH). The insets qualitatively display the main solutions: filled
  (open) circle for the stable (unstable) fixed point, and solid
  (dashed) curves for the stable (unstable) limit cycle.  The GH point
  is located at GH$=(309.5748,0.043262)$, $\beta=0$, and other
  parameters are as in (a).  (c) Bifurcation analysis in
  $\beta-\gamma$ space. We show three supercritical Hopf curves
  ($\alpha=50,100,200$). The area with sustained oscillations is found
  below the corresponding curve.  Other parameters are as in (a).}
 \label{alphabeta1}%
\end{figure*}

To measure how the stereoselectivity of polymerization and
depolymerization influence the appearance of oscillations, we perform
a bifurcation analysis in $\beta-\alpha$ space
(Fig.~\ref{alphabeta1}(a)). Since $c=1$ and other parameters are also
as in Fig.~\ref{calpha1}, for $\beta=0$, stable oscillations are found
for $\alpha > \alpha_H$ (at least for $\alpha < 1500$). If $0<\beta <
0.020445$, i.e., for only weak heterochiral depolymerization, the
range of $\alpha$ that allows for stable oscillations shrinks, with a
lower bound that increases and an upper bound that decreases. If
$\beta >0.020445$, there is no value of $\alpha$ that admits stable
oscillations. At the same time, from the figure we deduce that the
range of allowed values of $\beta$ is maximal for $\alpha=102.3$. Also
in this parameter space, we observe a Generalized Hopf point at
GH$=(358.033236,0.011450)$. To the left of that point, the Hopf
bifurcation is supercritial, to the right it is subcritical. There is
a LPC curve emerging at the GH point that in this scenario lies above
the subcritical Hopf curve. This implies that the local bifurcation
scenario is slightly different to the one displayed in
Fig.~\ref{calpha1}. The insets show the main solutions.  In
particular, we have a stable fixed point coinciding with a stable
limit cycle between the subcritical Hopf curve and the LPC curve. This
represents a bistability of oscillations with a stationary state.
However, this scenario is not complete. As we know from
Fig.~\ref{calpha1}, for $\beta=0$ and $\alpha>\alpha_H$, there exists
also another unstable limit cycle that, however, does not undergo any
bifurcation in this parameter range (and hence is not added to the
insets).

To check the relative impact of the stereoselectivity of
polymerization and epimerization, in Fig.~\ref{alphabeta1}(b), we
explore the region of stable oscillations in $\gamma-\alpha$ space.
Again, the other parameters are as in Fig.~\ref{calpha1}, implying
that for $\gamma=0$, stable oscillations are found for $\alpha >
\alpha_H=38.24$.  The functional form of the Hopf and LPC curves and
the location of GH point are qualitatively similar to the scenario in
the $\beta-\alpha$ space. Therefore, if $0<\gamma < 0.083136$, i.e.,
for only weak heterochiral epimerization, the range of $\alpha$ that
allows for stable oscillations shrinks, with a lower bound that
increases and an upper bound that decreases. If $\gamma >0.083136$,
there is no value of $\alpha$ that admits stable oscillations. At the
same time, from the figure we deduce that the range of allowed values
of $\gamma$ is maximal for $\alpha=90.05$.
\begin{figure*}
 \centering 
\includegraphics[width=16cm]{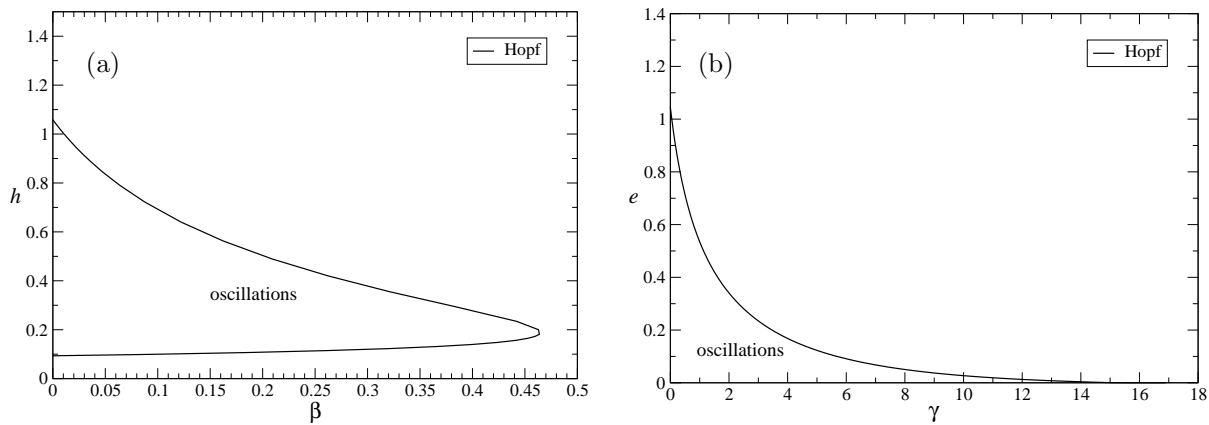}
\caption{(a) Bifurcation analysis in $h-\beta$ space. We show a
  supercritical Hopf curve for the parameters $\alpha=50$,
  $a=c=p=e=1$, and $b=\gamma=0$. The area with sustained oscillations
  is found to the left of the curve. (b) Bifurcation analysis in
  $e-\gamma$ space. We show a supercritical Hopf curve for
  $\alpha=50$, $a=c=p=h=1$, and $b=\beta=0$.  The area with sustained
  oscillations is found below the curve.}
 \label{hbeta1}%
\end{figure*}

To complete the view on relative impact of heterochiral
polymerization, depolymerization, and epimerization on oscillations,
in Fig.~\ref{alphabeta1}(c), we explore the $\beta-\gamma$ space for
fixed $\alpha$. As can be already inferred from the above figures,
there is a minimum value for $\alpha=\alpha_H=38.24$, below which no
combination of $(\gamma,\beta)$ yields stable oscillations. However,
if $\alpha>\alpha_H$, small values of $\gamma$ and $\beta$ are
allowed. As we increase $\alpha$, this area first increases and
decreases later, once the relative maxima of the $\beta-\alpha$ and
$\gamma-\alpha$ curves are passed.

In an analogous way as our study of $p$ and $\alpha$, we can compare
the influence of the depolymerization rate $h$ vs. its
stereoselectivity $\beta$ (Fig.~\ref{hbeta1}(a)). We choose
$\alpha=50$ that for $\beta=0$ and $h=1$ (and other parameters as
above) corresponds to stable oscillations. We find that while for
$h=1$ the allowed interval of $\beta$ is small, we see that if we
decrease $h$, the range of $\beta$ values giving rise to oscillations
becomes larger. This means that the depolymerization process needs not
to be very unefficient for heterochirals in order to produce
oscillations.  There exists an optimal value $h \approx 0.2$ where the
favorable $\beta$ range becomes largest.  Nevertheless, there is a
minimum value $h=0.0931$ below which no stable limit cycles are found.

Similarly, we study in Fig.~\ref{hbeta1}(b) the influence of the
$\gamma$. Interestingly, we do not observe a minimal threshold for $e$
as for $h$. To the contrary, we find that for $e\to 0$, the range of
allowed values of $\gamma$ becomes maximal, and reaches values close
to $14$. However, we should have in mind that if $e=0$, the value of
$\gamma$ becomes irrelevant since $\gamma e=0$. Furthermore, since for
$e=0$ the Jacobian matrix becomes singular, a more detailed analysis
should be performed to investigate the bifurcation scenario in that
limit. We remark that stable oscillations are also possible for
$\gamma=1$, i.e., the absence of any stereoselectivity.

\begin{figure}[!h] 
 \centering 
\vspace{0.6cm}
\includegraphics[width=8cm]{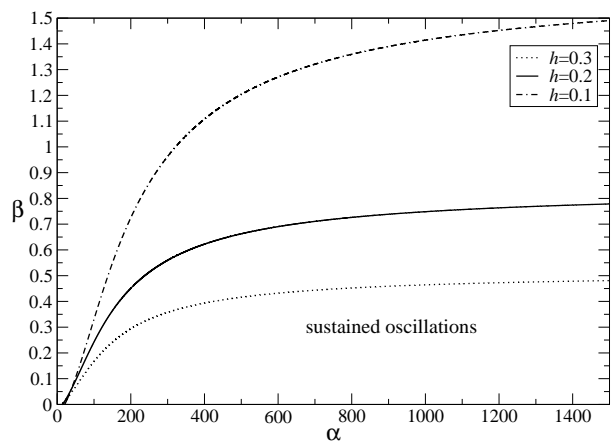}
\caption{Bifurcation analysis in $\beta-\alpha$ space. We show three
  supercritical Hopf curves ($h=0.3,0.2,0.1$). The area with sustained
  oscillations is found below the corresponding curve.  Other
  parameters are $a=c=\gamma=1$, $e=0.1$, $p=0.3$, and $b=0$.}
 \label{alphabetaN2}%
\end{figure}

To complete our study, we revisit the $\beta-\alpha$ space, but with a
parameter set for which $\gamma=1$ (Fig.~\ref{alphabetaN2}). The
motivation is to find stable oscillations in absence of any
stereoselective de- and epimerization.  We choose $c=1$, $e=0.1$, and
$p=0.3$, and consider three values for $h$. We see that the region of
oscillations increase for smaller $h$, and in particular, for $h=0.1$
and $\beta=1$, we have a critical value for $\alpha=321.447$ above
which a stable limit cycle is found. This is just an example to
demostrate that $\beta\ll 1$ and $\gamma\ll 1$ are not necessary
conditions for the presence of sustained oscillations.

\section{Conclusions}
\label{sec:disc}

In this paper, we have studied a simple polymerization model
(actually, dimerization), aimed originally to explain the emergence of
chiral states, in the context of the onset of {\emph{chiral
    oscillations}}. These oscillations represent temporal changes in
the enantiomeric excess of the underlying system, i.e., the continuous
change of a state, where one of the chiral states momentarily
dominates, into a subsequent state where the opposite handed state is
more abundant.

During decades, models for explaining the emergence and amplification
homochirality have been focused on the creation of stable stationary
states, typically associated with chiral states of maximum
enantiomeric excess. The description of temporally oscillatory states
in these models and the discussion of what these chiral oscillations
mean, are rare. 
Chiral oscillatory
states have been reported by Iwamoto in two different systems
motivated by traditional models for chemical
oscillations~\cite{IwamotoPCCP02,IwamotoPCCP03} and shortly after for
the APED model~\cite{PlassonPNAS04} studied here. Nevertheless, chiral
oscillations were not studied in detail in the latter paper, and no
explanation was given nor relevance is attributed to this effect.
Here, we have studied in detail the onset of chiral oscillations, and
have demonstrated that they are associated with temporal oscillations
of the underlying chemical species, i.e., chemical
oscillations. Mathematically speaking, the onset of oscillations is
described by a supercritical Hopf bifurcation scenario, implying that
oscillations have small amplitude and finite frequency (opposed to
other mechanisms where oscillations arise with finite amplitude and
vanishing frequency). The Hopf bifurcation is of codimension-1, and
oscillations are found in large parameter regions and hence represent
a generic solution of the system.

While it has been suggested~\cite{PlassonPNAS04} that the reason for
observing chiral oscillations is that heterochiral polymerization is
favoured over homochiral polymerization, we test this hypothesis and
clarify in more detail what are the conditions that enable the
stabilization of chiral oscillations within the framework of the
rather general APED model. We show continuation results varying
$\alpha$, $\beta$, $\gamma$, $p$, $h$, $e$, and $c$ (keeping $a=1$,
$b=0$), demonstrating that oscillations are in fact a typical solution
of the model.  In particular, for the $\alpha-\beta-\gamma$ space, we
measure the impact of hetero- vs.  homochiral polymerization,
hydrolysis, and epimerization, and find that $\alpha \gg 1$,
$\beta<1$, and $\gamma<1$ are favorable, in particular $\alpha> 38.24$
for $\beta=0$, and $\gamma=0$, where $c=1$ is the total mass of the
system and other the parameters are also equal to unity. This means
that oscillations are favored for heterochiral polymerization and
homochiral de-polymerization and epimerization. This is a noteworthy
result, especially in light of recent analysis of
Ref.~\cite{CrusatsCPC09} using the Frank scheme to model the Soai
reaction, which argued that an ideal system for kinetically controlled
spontaneous mirror symmetry breaking should consist of a fast
exergonic mutual inhibition step leading to the formation of the
heterodimer LD (i.e., $\alpha \gg 1$) and which must be more exergonic
that the possible homodimerization leading to LL and DD. The latter
condition translates into the constraint $\alpha \gg \beta$ in terms
of the APED model.  In all cases observed, the oscillations arise from
a achiral state.  Oscillations are not occurring for any amout of
total mass in the system. We observe that there is a lower bound (even
before the racemic solution reaches the dead state) and also an upper
bound.

We show stable sustained oscillations for a wide range of parameter
sets. In particular, $\beta$ and $\gamma$ can also be chosen to unity,
i.e., depolymerization and epimerization need not be stereoselective
in order to produce oscillations, although heterodimers must be
polymerized preferentially. Recent experimental results demonstrate
the stereoselectivity of polymerization reactions and show in detail
that the stereoselectivities can be modified experimentally (e.g., by
pH) in magnitude and can be found to be both larger or smaller than
unity~\cite{DangerAb10,PlassonOLEB11}.

The presence of oscillations of chirality adds a further element of
uncertainty to the determination of the sign of the chirality that is
finally transmitted to the system (say, within a closed reacting
system with necessarily damped chiral oscillations).  In some
parameter regions, even bistability of oscillations with the racemic
state are observed, implying that then it is a matter of the initial
condition which state is realized.

Until recently, chiral oscillations have not been reported in
experiments, and hence the possibility for studying their onset in
theoretical models has received sparse attention.  However, there is
now growing interest in oscillatory phenomena close to the onset of
chirality, as chiral oscillations have been reported in different
experimental systems consisting of carboxyclic acids (profens, amino
acids, hydroxy acids)~\cite{SajewiczJPOC10,SajewiczJSC10}. There,
several theoretical models are presented and checked whether they can
account for the experimental observations. Among them, the APED model
studied here, is the only one that was already known to display chiral
oscillations, while the other
ones~\cite{HyverJCP85,BykovCES87,PeacockLopezBpC97} had to be modified
or reinterpreted in order to account for chiral, rather than just mere
chemical oscillations. Here, we do not aim to discuss any of those
models with regard to the experiments (the reader is referred
to~\cite{SajewiczJPOC10,SajewiczJSC10} directly). 

For the Soai reaction, Buhse presented a simplified
model~\cite{BuhseTA03} that was later modified to investigate temporal
oscillations in continuous flow and semibatch
conditions~\cite{MicskeiJPCB08}. While there are stable oscillations
of the reactants in a continuous flow scenario, they are apparently
not associated with oscillations in the enantiomeric excess. This
means that chemical oscillations are not related to chiral
oscillations, as they are in the APED model. On the other hand, in the
case of a semibatch system, oscillations necessarily fade out and give
rise to a chiral state, but transient oscillations in the enantiomeric
excess can be observed.

\section*{Acknowledgments} 
M.S. acknowledges financial support from MICINN (currently
MINECO) through project FIS2011-27569 and from the Comunidad
Aut\'onoma de Madrid, project MODELICO (S2009/ESP-1691) and
acknowledges very useful help from W. Govaerts and V. De Witte using
MatCont and useful discussions with R. Plasson. 
The research of C.B. and D.H. is supported by the Grant
AYA2009-13920-C02-01 from MICINN (currently MINECO) and C.B. has a
Calvo-Rod\'{e}s predoctoral contract from INTA.

\providecommand*{\mcitethebibliography}{\thebibliography}
\csname @ifundefined\endcsname{endmcitethebibliography}
{\let\endmcitethebibliography\endthebibliography}{}

\end{document}